

\input phyzzx

\hsize 6.5truein
\vsize 9.0truein
\hoffset 0.2truein
\voffset 0.06truein

\def\la{\mathrel{\mathpalette\fun <}}
\def\ga{\mathrel{\mathpalette\fun >}}
\def\fun#1#2{\lower3.6pt\vbox{\baselineskip0pt\lineskip.9pt
  \ialign{$\mathsurround=0pt#1\hfil##\hfil$\crcr#2\crcr\sim\crcr}}}
\def\md{\mu}
\def\mpl{m_{\rm Planck}}

\def\gf{G_{\rm Fermi}}
\def\singlespace{\baselineskip 12pt}

\def\medspace{\baselineskip 18pt}

\def\doublespace{\baselineskip 24pt}

\def\solar{_\odot}

\def\sqr#1#2{{\vcenter{\vbox{\hrule height.#2pt
       \hbox{\vrule width.#2pt height#1pt \kern#1pt
          \vrule width.#2pt}
       \hrule height.#2pt}}}}

\doublespace
\PhysRevtrue
\nopubblock

\REF\COBE{G. F. Smoot, {\it et al.} Astrophys. J.  Lett. {\bf 396},
L1 (1992).}
\REF\LAMBDA{For a recent review of CDM, see
M. Davis, G. Efstathiou, C. S. Frenk, and S. D. M. White, Nature
{\bf 356}, 489 (1992).}
\REF\APM{G. Efstathiou, W. J. Sutherland, and S. J. Maddox, Nature {\bf 348},
705 (1990).}
\REF\GELB{M. Davis and P. J. E. Peebles, Astrophys. J. {\bf 267}, 465 (1983);
M. Davis, G. Efstathiou, C. S. Frenk, and S. D. M. White, Astrophys. J. {\bf
292}, 371 (1985); J. M. Gelb, Ph.D. Thesis, M.I.T. (1992).}
\REF\HYBRID{Q. Shafi and F. W. Stecker, Phys. Rev. Lett., {\bf 53},
1292 (1984);
R. K. Schaefer, Q. Shafi, and F. W. Stecker, Astrophys. J. {\bf 347},
575 (1989); J. A. Holtzman, Astrophys. J. Suppl. {\bf 71}, 1 (1989);
A. Klypin, J. Holtzman, J. Primack, and E. Reg\"os, UC Santa Cruz preprint
SCIPP 92/52.}
\REF\WRIGHT{E. Wright, {\it et al.} Astrophys. J.  Lett. {\bf 396},
L13 (1992).}
\REF\US{For a paradigm explaining why $\Omega_{\rm CDM} \sim
\Omega_{\rm baryon}$, see
S. Dodelson and L. M. Widrow, Phys. Rev. Lett. {\bf 64}, 340 (1990).}
\REF\SHAFI{See, however Ref. 5 as well as J. Madsen, Phys. Rev. Lett. {\bf 69},
571 (1992) and N. Kaiser, R. A. Malaney, and G. D. Starkman, `Neutrino-Lasing
in the Early Universe', CITA preprint (1993).}
\REF\BOND{J. R. Bond, A. S. Szalay, and M. S. Turner, Phys. Rev. Lett. {\bf
48}, 1636 (1982).}
\REF\LEFTHAND{Of course one could imagine Majorana mass tems for
the left handed neutrinos as well but these are {\it not}
$SU(2)_L\times U(1)_Y$
invariant and hence involve new physics [e.g. a Higgs triplet, see
G. B. Gelmini and M. Roncandelli, Phys. Lett. B {\bf 99}, 411 (1981)].}
\REF\LANGACKER{P. Langacker, University of Pennsylvania Report No.
UPR 0401T, 1989 (unpublished).}
\REF\DOLGOV{A. Dolgov, Sov. J. Nucl. Phys. {\bf 33}, 700 (1981).}
\REF\MANOHAR{A. Manohar, Phys. Lett. B {\bf 186}, 370 (1987).}
\REF\BARDOL{R. Barbieri and A. Dolgov, Phys. Lett. B {\bf 237},
440 (1990); Nucl. Phys. {\bf B349}, 742 (1991).}
\REF\KIMMO{K. Enqvist, K. Kainulainen, and J. Maalampi, Phys. Lett. B
{\bf 244}, 186 (1990); {\bf 249}, 531 (1990); K. Enqvist, K. Kainulainen,
and M. Thomson, Nucl. Phys. {\bf B373}, 498 (1992).}
\REF\CLINE{J. M. Cline, Phys. Rev. Lett. {\bf 68}, 3137 (1992).}
\REF\NOTNUE{These expressions hold for $\nu_\mu$ or $\nu_\tau$; the
interaction rate and mixing angle in matter differ somewhat for $\nu_e$.}
\REF\REFBBN{For similar limits in a slightly different language, see
Refs. \LANGACKER,\BARDOL,\KIMMO,\CLINE.}
\REF\WALKER{T. P. Walker {\it et al.}, Astrophys. J. {\bf 376}, 51 (1991).}
\REF\TG{S. Tremaine and J. Gunn, Phys. Rev. Lett. {\bf 42}, 407 (1979).}

\titlepage
\singlespace
\rightline{\hfill March 1993}

\title{Sterile Neutrinos as Dark Matter}
\author {Scott Dodelson$^{1,}$\footnote{\sharp}{E-mail address:
dodelson@fnal.fnal.gov} and
Lawrence M. Widrow$^{2,}$\footnote{\natural}{E-mail
address: widrow@orca.cita.utoronto.ca}}

\address {$^1$NASA/Fermilab Astrophysics Center\break
          Fermi National Accelerator Laboratory\break
          P.O. Box 500, Batavia, IL 60510 \break
                    \break
          $^2$Department of Physics,
           Queen's University\break
        Kingston, Canada, K7L 3N6\break
        and\break
        Canadian Institute for Theoretical Astrophysics\break
        University of Toronto, Toronto, Canada, M5T 1A7\break}

\vskip 0.1in
\centerline{ABSTRACT}
\vskip 0.05in

The simplest model that can accomodate a viable nonbaryonic dark matter
candidate is the standard electroweak theory with the addition of right-handed
or sterile neutrinos. This model has been studied extensively in
the context of the hot dark matter scenario. We reexamine this model and
find that hot, warm, and cold dark matter are all possibilities.  We
focus on the case where sterile neutrinos are the dark matter.  Since
their only direct coupling is to left-handed or active neutrinos, the
most efficient production mechanism is via neutrino oscillations. If the
production rate is always less than the expansion rate, then these
neutrinos will never be in thermal equilibrium. However, they may still play a
significant role in the dynamics of the Universe and possibly
provide the missing mass necessary for closure. We consider a single
generation of
neutrino fields $\left (\nu_L,\,\nu_R\right )$ with a Dirac mass, $\md$,
and a Majorana mass for the right-handed components only, $M$.  For
$M\gg \md$ we show that the number density of sterile neutrinos is
proportional to $\md^2/M$ so that the energy density today is
{\it independent of} $M$.  However $M$ is crucial in determining the large
scale structure of the Universe.  In particular,
$M\simeq 0.1-1.0 {\rm ~keV}$ leads to warm
dark matter and a structure formation scenario that may
have some advantages over both the standard hot and cold dark matter scenarios.
\endpage

\medspace
\FIG\FIGONE{Figure 1. Mass scales in hot dark matter and warm dark matter
as a function of scale factor. $M_H$ (solid line) gives the mass within
the horizon. Long dashed lines give the free streaming mass for a $30$
eV ($M_{FS,30}$) and 300 eV ($M_{FS,300}$) neutrino. Short dashed lines
are the Jeans mass for a $30$
eV ($M_{FS,30}$) and 300 eV ($M_{FS,300}$) neutrino.}
The COBE DMR experiment's recent detection of large-scale anisotropy in the
cosmic microwave background$^\COBE$
has considerably strengthened the view that the
large scale structures seen today evolved from very small primeval density
inhomogeneities.  Still, the two primary ingredients which
dictate how structure
forms, namely the nature of dark matter and the shape of the
primeval fluctuation spectrum, remain unknown.

The best studied and perhaps most successful model for structure formation is
known as the cold dark matter (CDM) theory$^\LAMBDA$.  In the standard CDM
model, the
Universe is assumed to be spatially flat ($\Omega=1$) with $90-95\%$ of the
mass density in dark matter and the balance in baryons ($5-10\%$) and
photons and light neutrinos ($\ll 1\%$).  Primeval fluctuations are generated
during inflation and are Gaussian with a scale-invariant spectrum.  CDM,
with the additional assumption that galaxy formation is `biased'
to occur first at the highest peaks in the density fluctuation spectrum can
successfully explain galaxy-galaxy and cluster-cluster correlation
functions on scales of order $1-5$ Mpc and is at least consistent with the
morphology of galaxies.
However, CDM now appears to be inconsistent with various sets of observational
data.  Perhaps its greatest difficulties come with large scale structure
data such as the APM galaxy survey$^\APM$,
which suggest more power on large scales than standard CDM model
predictions.  On small scales, the observed pairwise velocity
dispersion for galaxies
appears to be significantly smaller than those predicted by CDM$^{\GELB}$.

One alternative$^{\HYBRID}$
 which has recently received a fair bit of attention is
cold + hot dark matter (C+HDM).  HDM is usually taken to be a light neutrino
with $m_\nu=\left (92\Omega_\nu h^2\right ){\rm eV}$
where $H=100h {\rm ~km/sec/Mpc}$ is the Hubble parameter.
In models with HDM alone, the processed fluctuation spectrum is characterized
by the typical distance a neutrino travels over the history of the Universe,
$\lambda_\nu\simeq 40\left (30~{\rm eV}/m_\nu\right ) {\rm Mpc}$.  This in turn
sets the mass scale below which damping occurs due to free-streaming,
$M_{\rm FS} \equiv 4\pi\rho\left (\lambda_\nu/2\right )^3/3\simeq
 3\times 10^{15} \left (30\,{\rm eV}/m_\nu\right )^2 \Omega_\nu^{-1}\,
M_{\solar}$. In HDM
models, the first structures to form are pancake-shaped objects of size
$\lambda_\nu$
with smaller scale structures such as galaxies and clusters forming later via
fragmentation.  However, we know from the galaxy correlation function,
that the scale which is just becoming nonlinear today is
around $5h^{-1} {\rm Mpc}$.
Essentially, the problem with HDM alone is that $\lambda_\nu$ is too large:
If galaxy formation occurs early enough to be consistent with high-redshift
galaxies and quasars, then structures on $5h^{-1} {\rm Mpc}$ will overdevelop.
The hope is that C+HDM will combine the successes of both models.  In
fact, a survey$^{\WRIGHT}$ of models with various amounts of
hot dark matter, cold dark matter and baryons points to
$\Omega_{\rm baryon}=0.1,~\Omega_{\rm CDM}=0.6,~\Omega_\nu=0.3$
and a Hubble constant of $h=0.5$ as the best fit model for
microwave anisotropy data, large scale structure
surveys, and measures of the bulk flow within a few hundred megaparsecs.

As appealing as C+HDM may be for large scale structure phenomenology,
it is somewhat unpalatable from the point of view of particle physics.
Since there are no stable, neutral, massive
particles in the `standard model' for electroweak interactions, the existence
of nonbaryonic dark matter implies new physics.  Given that the existence of
the baryon-antibaryon asymmetry also requires new (and probably distinct)
physics, it seems already a great coincidence that $\Omega_{\rm DM}$
and $\Omega_{\rm baryon}$ be as close as they are$^{\US}$.
Two types of dark matter
imply further additions to the standard model with yet another coincidence in
order to
have $\Omega_{\rm HDM}$, $\Omega_{\rm CDM}$, and $\Omega_{\rm baryon}$ all
within one or two orders of magnitude of each other$^{\SHAFI}$.

By far the simplest dark matter candidate, at least
from the point of view of particle physics is the neutrino.
Massive neutrinos require
only the addition of right-handed or sterile neutrino
fields to the standard model.
In fact, it is the {\it absence} of right-handed neutrinos
that seems contrived in
light of the fact that all other fermions in the standard model have both left
and right-handed components.

Here we focus on the possibility that
sterile neutrinos are the
dark matter and that they are somewhat heavier but less abundant than the
usual HDM neutrino.
Such a `warm' dark matter particle may have advantages for structure
formation over both hot and cold dark matter scenarios.
Our work is similar in some respects to that of Bond, Szalay, and
Turner$^{\BOND}$ who consider a
particle that is in thermodynamic equilibrium at early times but
decouples before ordinary neutrinos do so that
$g_*$, the number of effectively massless degrees
of freedom, is relatively high $(g_*\ga 100 )$.
Warm dark matter has been for the most part been ignored, to a large extent
because there have been no compelling candidates proposed thusfar.  In part,
the motivation for this work is to propose a `realistic' warm dark matter
candidate.

For simplicity, we consider only one generation of neutrinos.  The mass terms
for the neutrinos are then$^{\LEFTHAND}$:
$${\cal L}=\mu\left ({\phi\over v}\right )\bar\nu_L\nu_R+M\nu_R\nu_R
+{\rm~ h.c.}\eqn\LAGR $$
where $\phi$ is the standard model Higgs field with $\langle\phi\rangle=v$.
The usual HDM case, wherein the active neutrinos constitute the dark matter,
corresponds to $\left\{\mu=92h^2{\rm eV},~ M\ll \mu\right\}$
or $\left\{\mu^2/M = 92h^2{\rm eV},~ M\gg\mu\right\}$.
When sterile neutrinos are the dark matter, the relevant mass is $M$.
At tree-level, $\nu_R$ couples only to $\nu_L$ and
therefore the
most efficient way to produce sterile
neutrinos$^{\LANGACKER,\DOLGOV,\MANOHAR}$
is via oscillations
$\nu_L \rightarrow \nu_R$.  The probability of observing a right-handed
neutrino after a time $t$ given that one starts with a pure
monoenergetic left-handed
neutrino is ~$\sin^2{2\theta_M} \sin^2{vt/L}$~ where $\theta_M$ is the `mixing
angle', $L$ is the oscillation length, and $v$ is the velocity of the
neutrinos.  In vacuum, and with $\mu\ll M$ (see-saw model)
$\theta_M=\md/M$ and $L=4E/\left (M^2-\md^2\right )$
where $E$ is the energy of the neutrinos.
In the early Universe, the observation time
$t$ is replaced by the interaction time for the left-handed neutrinos.
Recent work$^{\BARDOL,\KIMMO,\CLINE}$
has fine-tuned this picture taking into account the effect of
finite density and temperature on the mixing angle.

Here we are interested in the case where the right-handed neutrinos are
produced at temperatures of order $100~{\rm MeV}$
though the production rate is never so fast that they equilibrate.  We begin
with the Boltzmann equation for the sterile neutrinos:
$$\left (
 {\partial \over \partial t} ~-~ H E{\partial \over
\partial E}
\right ) f_S(E,t) ~=~ \left [{1\over 2}\sin^2(2\theta_M(E,t))~\Gamma(E,t)
\right ]f_A(E,t) \eqn\BOLTZ $$
where $f_S$ and $f_A$ are the distribution functions of the sterile
and active neutrinos.  In the epoch under consideration $\left (
 T \gg 1 {\rm ~MeV}\right )$ the left-handed neutrinos are in
thermal equilibrium so that $f_A=\left (e^{E/T}+1\right )^{-1}\simeq
\left (e^{p/T}+1\right )^{-1}$.  The quantity
in square brackets is the probability per time of an active neutrino
converting into a sterile one$^{\CLINE}$
where we have used the fact that for
parameters of interest, the collision time is always much greater than
the oscillation time (i.e. $\sin^2{vt/L}$ averages to $1/2$).
The mixing angle and the collision rate are$^{\NOTNUE}$
$$ \sin^2(2\theta_M) = {\md^2 \over \md^2 +
[(c\Gamma E/M) + (M/2)]^2} \ \ \  ;
 \ \ \ \Gamma \simeq {7\pi\over 24} \gf^2 T^4 E \eqn\MIXING $$
where
$c \simeq 4 \sin^2(2\theta_W) /15\alpha\simeq 26.$

To get a feel for when and how many sterile neutrinos are produced, we derive
the equation for $r\equiv n_S/n_A$ where $n_i\equiv 2\int d^3p
f_i/\left (2\pi\right )^3$ is the number density of sterile (active) neutrinos
with $i=s\,\left (i=A\right )$.  Changing the time variable from $t$ to $a$,
the Robertson-Walker scale factor
and integrating Eq. \BOLTZ\ over momenta, one finds that
$${dr\over d\ln{a}}={\gamma\over H} +r{d\ln{g_*}\over d\ln{a}}\eqn\RATIO $$
where
$$\gamma\equiv {1\over n_A}\int {d^3p\over \left (2\pi\right )^3}\,
\sin^2{2\theta_M(p,T)}\,\Gamma(p,T)\, {1\over e^{p/T}+1}~,
\eqn\DEFGAM $$
and we have used the fact that $g_*a^3T^3 = {\rm constant}$.
For $g_*$ constant, $\gamma/H$
gives the number of sterile neutrinos, relative to the number of active
neutrinos, that are produced in each log-interval of $T$.  Substituting Eq.
\MIXING, using $H = 1.66g_*^{1/2} T^2/\mpl$,
and taking the limit $M\gg \md$, we find that
$${\gamma\over H}={13\over g_*^{1/2}}\left (\mu\over {\rm eV}\right )^2
\left ({\rm keV}\over M\right )
x\int_0^\infty{y^3dy\over\left
(e^y+1\right )\left (1+x^2y^2\right )^2} \eqn\GAMOVH $$
where $x\equiv 78\left (T/{\rm GeV}\right )^3\left ({\rm keV}/M\right )$.
Taking $g_*=10.8$ and doing the integral numerically, we
find that $\gamma/H$ reaches a peak value of $1.9\left (\mu/{\rm eV}\right
)^2\left ({\rm keV}/M\right )$ when $x\simeq 0.19$ or
$T=T_{\rm max}\simeq 133
\left (M/{\rm keV}\right )^{1/3} {\rm ~MeV}$ and falls off as
$T^3$ for $T\ll T_{\rm max}$ and $T^{-9}$ for $T\gg T_{\rm max}$.
Evidently, the number density in sterile neutrinos is proportional to $M^{-1}$
so that the energy density is {\it independent of} $M$.  Note also that most of
the neutrinos are produced when the Universe has a temperature $T\simeq T_{\rm
max}$.  As will be discussed below, our calculations simplify if we can assume
that $g_*$ is constant.  Since $g_*$ changes abruptly at $T\simeq 200 {\rm~
MeV}$ and varies slowly for $200 {\rm ~MeV}\ga T\ga 20 {\rm~ MeV}$, this
assumption will be pretty good for $M\la {\rm keV}$ but breakdown for masses
much larger than this.

Our interest is in the
structures which form in a $\nu_R$-dominated Universe and
we therefore require the full sterile neutrino distribution function.
Here, we make the
assumption that $g_*$ is constant.
Using $\partial f_S/\partial t=-HT\partial
f_S/\partial T$ and the identity
$$T\left ({\partial f_S\over\partial T}\right )_E+
E\left ({\partial f_S\over\partial E}\right )_T=
T \left ({\partial f_S\over \partial T}\right )_{E/T}\eqn\IDEN $$
and changing the integration variable from $T$ to $x$
one finds
$$ {f_S\over f_A} = {7.7 \over  g_*^{1/2}}\left ({\mu\over {\rm eV}}\right )^2
\left ({{\rm keV}\over M}\right )~y
        \int_x^\infty { dx'  \over
    \left (1+y^2x'^2\right )^2 }\eqn\ANALYTIC $$
where $y \equiv E/T$.
In general, the right hand side of Eq. \ANALYTIC\ is a complicated function
of $E$ and therefore will have a different energy dependence than $f_A$.
There is no reason to expect otherwise: high energy and low energy
neutrinos oscillate at different rates.  Moreover, these rates change with
temperature.
HOWEVER, for $T\ll T_{\rm max}$ the lower limit of the integral can be set to
zero and the right hand side of \ANALYTIC\
becomes independent of $E$ and $T$.
In this limit, the integral is easily done and we find
$$ f_S = \left (6.0/g_*^{1/2} \right)\left ( \mu/{\rm~ eV}\right)^2
\left ({\rm keV}/M \right )~ f_A .\eqn\ANTWO$$
$f_S$ has the same functional form as $f_A$ and therefore
$\Omega_S/\Omega_\nu = \left (M/m_\nu\right )\left (f_S/f_A\right )$.
{}From the relation $m_\nu/\Omega_\nu\simeq 92 h^2{\rm ~eV}$ we find that
$\Omega_S=1$ for $\mu=0.22 h {\rm~ eV}$ where we have again set $g_*=10.8$.
Finally, we note that the contribution of sterile neutrinos to the energy
density of the Universe at the time of primordial
nucleosynthesis$^{\REFBBN}$ must be $\la
0.5$ times the contribution of a light neutrino species if
standard big bang nucleosynthesis$^{\WALKER}$
is to be believed.  This in turn implies
that $M \ga 200 h^2 {\rm eV}$;  that is, if sterile neutrinos are the dark
matter then they are necessarily more massive than the standard HDM.

How do perturbations evolve when a sterile neutrino species is the dark
matter? Several guiding principles help us understand
the processed power spectrum. First, structure within the horizon
grows only after
the dominant component of matter becomes nonrelativistic and therefore the size
of the horizon at matter-radiation equality $\lambda_H(a=a_{eq}) \equiv
a_{eq}\int_0^{a_{eq}} dt'/a(t')$, defines a characteristic scale.
Second, perturbations on scales smaller than the Jeans length
$\lambda_J \equiv (\pi v_s^2 m_{\rm Planck}^2/\rho)^{1/2}$ (where $v_s$ is the
speed of sound) oscillate like pressure waves. Finally, for
neutrinos, or any particle which is not completely
non-relativistic,  perturbations on scales smaller
than the free streaming scale
$\lambda_{FS} \equiv a\int_0^t dt' \langle (p/E)^2 \rangle^{1/2} /a(t')$
are exponentially damped.
With the distribution function in Eq. \ANTWO, one can calculate these
scales for sterile neutrinos.
Figure \FIGONE\ shows the relevant mass scales $(=4\pi \rho (\lambda/2)^3/3)$
as a function of the
scale factor for the sterile neutrinos discussed here and for
an ordinary light neutrino dark matter candidate. For light neutrinos,
the damping scale and the horizon scale at equality are roughly equal
$[\sim 10^{15} M_{\solar}]$, of order supercluster size.
This scale is the first to go non-linear. For sterile neutrinos, there
is a large disparity between the two characteristic scales, so that
perturbations with $10^{13} M_{\solar} \la M \la 10^{15} M_{\solar}$
are processed similarly; given an initial Harrison-Zel'dovich
spectrum, they should all the
the same final amplitude in linear theory. Power on scales smaller
than this should be completely damped.


In conclusion, we have proposed a candidate for warm dark matter that
exists in the simplest extension of the standard model. Warm dark matter
has several advantanges over cold or hot dark matter, resulting
from the fact that the power on
scales of order $1-5$ Mpc is less than in CDM but greater than in HDM.
In particular, the pairwise velocity dispersions in a WDM
universe are likely to be smaller than in CDM and hence more in accord
with observations. Since there is more power on small scales than in HDM,
the epoch of galaxy formation is likely to be earlier and hence the observed
high redshift quasars pose less of a problem for this model than for HDM.
On large scales there is little difference among the three models: at
present they all seem to be incompatible with the APM survey.
Another advantage WDM has over HDM is that since the neutrino mass
is higher, it is possible to fit more neutrinos into a given galaxy,
thus evading Tremaine-Gunn limits$^{\TG}$. Finally
we point out a unique signature of WDM is an increase in the
predicted primordial helium abundance; since a neutrino species that is in
thermal equilibrium at the time of big bang nucleosynthesis
adds $\Delta Y = 0.012$ to the primordial helium mass fraction,
sterile neutrinos add
$$ \Delta Y = .01 \left( {100 h^2 {\rm eV} \over M} \right), \eqn\DELY $$
a potentially detectable deviation from the standard prediction.
\bigskip
\noindent

ACKNOWLEDGEMENTS

It is a pleasure to thank David Spergel for helpful comments.
The work of SD was supported in part by the DOE and NASA
grant NAGW-2381 at Fermilab.

\refout

\figout

\bye